\documentclass[showpacs,preprintnumbers,amsmath,amssymb,floatfix]{revtex4}

\usepackage{graphicx}
\usepackage{epsfig}
\usepackage{amsfonts}
\usepackage{amssymb}
\usepackage{epsf}
\newcommand{\insertplot}[5]{\begin{figure}
 \hfill\hbox to 0.05in{\vbox to #5in{\vfill
 \inputplot{#1}{#4}{#5}}\hfill}
 \hfill\vspace{-.1in}
 \caption{#2}\label{#3}
 \end{figure}}
\newcommand{\inputplot}[3]{
 \special{ps: plotfile #1}
}

\newcounter{fig}   \newcommand{\lbfig}[1]{\refstepcounter{fig}
\label{#1} }

\begin{document}

\title{
Rotating Black Holes in Dilatonic Einstein-Gauss-Bonnet Theory}

\author{
{\bf Burkhard Kleihaus, Jutta Kunz}
}
\affiliation{
{Institut f\"ur Physik, Universit\"at Oldenburg,
D-26111 Oldenburg, Germany}
}
\author{ 
{\bf Eugen Radu}
}
\affiliation{
{School of Theoretical Physics -- DIAS, 10 Burlington
Road, Dublin 4, Ireland}\\
{Department of Computer Science,
National University of Ireland Maynooth,
Maynooth, Ireland}
}
\date{\today}
\pacs{04.70.-s, 04.70.Bw, 04.50.-h}
\begin{abstract}
We construct generalizations of the Kerr black holes
by including higher curvature corrections 
in the form of the Gauss-Bonnet density coupled to the dilaton.
We show that the domain of existence of these 
Einstein-Gauss-Bonnet-dilaton (EBGd) black holes is
bounded by the Kerr black holes,
the critical EGBd black holes, and the singular extremal EGBd solutions.
The angular momentum of the EGBd black holes can exceed the Kerr bound.
The EGBd black holes satisfy a generalised Smarr relation.
We also compare their innermost stable circular orbits with those of the
Kerr black holes and show the existence of differences which
might be observable in astrophysical systems.
\end{abstract}
\maketitle

\noindent{\textbf{~~~Introduction.--~}}
Black holes (BHs) represent some of the most fascinating objects in nature.
In the past decades, 
observational evidence for their existence has increased tremendously,
making BHs essential ingredients of modern astrophysics on all scales, 
from stellar binaries to galaxies and active galactic nuclei.
 
The existence of BHs is an important prediction of general relativity, 
with the astrophysically most relevant Kerr solution 
found almost half a century ago.
String theory, however, suggests the existence of higher curvature corrections 
to the Einstein equations,
which may lead to new qualitative features of the solutions.
One of the simplest four-dimensional models with higher-curvature terms 
is the so-called Einstein-Gauss-Bonnet-dilaton (EGBd) gravity, 
which is obtained by adding to the Einstein action 
the four-dimensional Euler density multiplied
by the dilaton exponent together with the dilaton kinetic term.
EGBd gravity has the attractive features
that the equations of motions are still of second order 
and that the theory is ghost-free.
Although the BHs in this theory cannot be found in analytical form,
the static solutions were extensively studied perturbatively 
\cite{Mignemi:1992nt}, \cite{Mignemi:1993ce},  
and numerically
\cite{Kanti:1995vq}, \cite{Torii:1996yi}, \cite{Alexeev:1996vs}.
These BHs contain classical non-trivial dilaton fields
and thus can evade the classical ``no-scalar-hair'' theorem.

However, astrophysical black holes are expected to be highly spinning,
and nearly extreme rotating black holes have been observed in the sky  
\cite{McClintock:2006xd}.
Rotating solutions in EGBd theory have not yet been  
constructed, except in lowest order perturbation
theory \cite{Pani:2009wy}.
Therefore, we here address the question on how the GBd term affects 
the properties of rotating BHs.
We construct the EGBd generalizations of the Kerr solution
within a nonperturbative approach 
by directly solving the field equations
with suitable boundary conditions.
We exhibit their domain of existence 
and study their geodesics.

\noindent{\textbf{~~~The model.--~}}
We consider the following low-energy effective action
for the heterotic string
\begin{eqnarray}  
S=\frac{1}{16 \pi}\int d^4x \sqrt{-g} \left[R - \frac{1}{2}
 (\partial_\mu \phi)^2
 + \alpha  e^{-\gamma \phi} R^2_{\rm GB}   \right],
\label{act}
\end{eqnarray} 
where 
$\phi$ is the dilaton field
with coupling constant $\gamma$, $\alpha $ is a numerical
coefficient given in terms of the Regge slope parameter,
and
$R^2_{\rm GB} = R_{\mu\nu\rho\sigma} R^{\mu\nu\rho\sigma}
- 4 R_{\mu\nu} R^{\mu\nu} + R^2$ 
is the GB correction. 
 
We employ the usual Lewis-Papapetrou ansatz \cite{wald} 
for a stationary, axially symmetric spacetime with
two Killing vector fields $\xi=\partial_t$,
$\eta=\partial_\varphi$. 
In terms of the spherical coordinates $r$ and $\theta$, 
the isotropic metric reads \cite{Kleihaus:2000kg}
\begin{equation}
ds^2 = g_{\mu\nu}dx^\mu dx^\nu= -fdt^2+\frac{m}{f}\left(dr^2+r^2 d\theta^2\right) 
       +\frac{l}{f} r^2 \sin^2\theta
          \left(d\varphi-\frac{\omega}{r}dt\right)^2,
\label{metric} 
\end{equation}
where $f$, $m$, $l$ and $\omega$ are functions of $r$ and $\theta$, only.
The event horizon of these stationary black holes resides at a surface
of constant radial coordinate $r=r_{\rm H}$,
and is characterized by the condition $f(r_{\rm H})=0$.
At the horizon we impose the boundary conditions
$f=m=l=0$, $\omega= \Omega_{\rm H} r_{\rm H}$,
and $\partial_r \phi = 0$, 
where $\Omega_{\rm H}$ is the horizon angular velocity.
The boundary conditions at infinity,
$f=m=l=1$, $\omega=0$, $\phi=0$, 
ensure that the solutions are asymptotically flat with a vanishing dilaton.
Axial symmetry and regularity impose
the boundary conditions on the symmetry axis ($\theta=0$),
$\partial_\theta f = \partial_\theta l = 
\partial_\theta m = \partial_\theta \omega = 0$,
$\partial_\theta \phi = 0$,
and, for solutions with parity reflection symmetry, agree with
the boundary conditions on the $\theta=\pi/2$-axis.
The absence of conical singularities implies also $m=l$ at $\theta=0$.

The mass $M$, the angular momentum $J=aM$,
and the dilaton charge $D$ are obtained from the asymptotic expansion via
\begin{equation}
f \rightarrow 1 - \frac{2 M}{r},~~
\omega \rightarrow \frac{2 J}{r^2},
~~
\phi \rightarrow - \frac{D}{r}.
\label{asym}
\end{equation}
Of interest are also the properties of the horizon.
The surface gravity $\kappa_{\rm sg}$ is obtained from \cite{wald}
$\kappa^2_{\rm sg} = -1/4 (D_\mu \chi_\nu)(D^\mu \chi^\nu) $
where the Killing vector
$\chi = \xi + \Omega_H \eta$ 
is orthogonal to and null on the horizon.
Expansion near the horizon 
in $\delta = (r-r_{\rm H})/r_{\rm H}$ 
yields to lowest order $f=\delta^2 f_2(\theta)$,
$m=\delta^2 m_2(\theta)$,
showing that the surface gravity 
is indeed constant on the horizon.
The Hawking temperature $T_{\rm H}$ of the black hole is
 \begin{eqnarray}
 T_{\rm H}=\frac{\kappa_{\rm sg}}{2\pi}=\frac{1}{2 \pi r_{\rm H}}\frac{f_2(\theta)}{\sqrt{m_2(\theta)}}. 
 \end{eqnarray}
The entropy of these BHs
can be written in Wald's form \cite{Wald:1993nt} 
as an integral over the event horizon
\begin{eqnarray}
\label{S-Noether} 
S=\frac{1}{4}\int_{\Sigma_h} d^{3}x 
\sqrt{h}(1+ 2\alpha e^{-\gamma \phi} \tilde R),
\end{eqnarray} 
where $h$ is the determinant of the induced metric on the horizon 
and $\tilde R$ is the event horizon curvature. 

The event horizon properties and the global charges 
are related through the Smarr mass formula
\begin{eqnarray}
\label{Smarr}
M=2 T_{\rm H} S+2\Omega_{\rm H} J-\frac{D}{2\gamma},
\end{eqnarray} 
analogous in form to the mass formula of hairy black holes
\cite{Kleihaus:2002tc}.
This Smarr relation is obtained by starting from the Komar expressions,
and making use of the equations of motion,
and the expansions at the horizon and at infinity.

\noindent{\textbf{~~~The solutions.--~}}
We solve the set of five second order  coupled non-linear 
 partial differential equations for the functions $f,l,m,\omega$ and $\phi$ numerically,
subject to the above boundary conditions, employing a compactified  
coordinate  $x = (r - r_{\rm H})/(1+r)$.
As initial guess we use the static BH solutions in EGBd theory
with a given $\alpha>0$.
By increasing $\Omega_{\rm H}$ from zero, we obtain rotating BH solutions,
whose mass $M$, dilaton charge $D$ and angular momentum $J$
are determined from their asymptotic behaviour
(see eq.~(\ref{asym})).
For all numerical calculations we choose the dilaton coupling
$\gamma=1$.

Let us first address the domain of existence of these EGBd BHs.
In Fig.~\ref{f-1} (left) we show the scaled horizon area $a_{\rm H}=A_{\rm H}/M^2$
as function of the scaled angular momentum $j=J/M^2$. The domain of existence
is indicated by the shaded area. 
The Kerr BHs are all mapped to a single curve,
with the set of Schwarzschild BHs mapped to the point
$a_{\rm H}=1$, $j=0$. For these Einstein BHs
the scaled entropy $s=S/M^2$ and area are proportional, $s=a_{\rm H}/4$.
We observe that the Einstein BHs form one of the boundaries
of the domain of existence. With respect to the 
scaled entropy $s$, the Kerr values form the lower boundary
in the range of existence of Kerr solutions, i.e., for $j\le 1$.

The left border of the domain of existence ($j=0$) is formed by the static EGBd BHs.
These are known to exist in the range $4 \le s \le 4.8$,
or in terms of the scaled area $a_H$, exhibited in Fig.~\ref{f-1} (right),
in the range $0.85 \le a_{\rm H} \le 1$ \cite{Kanti:1995vq}. 
In this range the scaled area $a_{\rm H}$ 
of the static EGBd BHs
increases monotonically with increasing mass to the 
Schwarzschild value $a_{\rm H}=1$, while the scaled entropy
decreases monotonically with increasing mass.
The critical static solution has
the smallest scaled area and the largest scaled entropy
possible for a static EGBd BH.
The occurrence of this bound is seen
in the horizon expansion of the dilaton field,
where it originates when a square root vanishes
as the critical horizon size is reached \cite{Kanti:1995vq}.

Analogous to the static case, such critical solutions
exist also for the rotating EGBd BHs.
These critical black holes form the upper boundary
of the domain of existence, when considered in terms of
the scaled entropy versus the scaled angular momentum,
as seen in  Fig.~\ref{f-1} (left).
Also shown in the figure are
families of EGBd BHs with fixed scaled horizon angular velocity
$\Omega_{\rm H} \alpha^{1/2}$, which all begin at their
respective critical BH solution, corresponding to the smallest
scaled angular momentum for this value of $\Omega_{\rm H} \alpha^{1/2}$.
But where do these curves of fixed $\Omega_{\rm H} \alpha^{1/2}$ end?

The family of Kerr BHs ends at the extremal Kerr solution,
which precisely saturates the Kerr bound for the scaled angular momentum,
$j_{\rm Kerr} \le 1$.
Interestingly, the EBGd BHs can exceed the Kerr bound.
Indeed, we find $j \le 1.02$
as highlighted in the inset of Fig.~\ref{f-1} (right).
We observe that the families of EGBd BHs 
with fixed scaled horizon angular velocity $\Omega_{\rm H} \alpha^{1/2}$
all end, when a respective singular extremal EGBd solution is reached.
These singular extremal EGBd solutions form the final part 
of the boundary of the domain of existence of the EGBd BHs,
as seen in the inset of Fig.~\ref{f-1} (right).

\begin{figure}[t!]
\lbfig{f-1}
\begin{center}
\includegraphics[height=.25\textheight, angle =0]{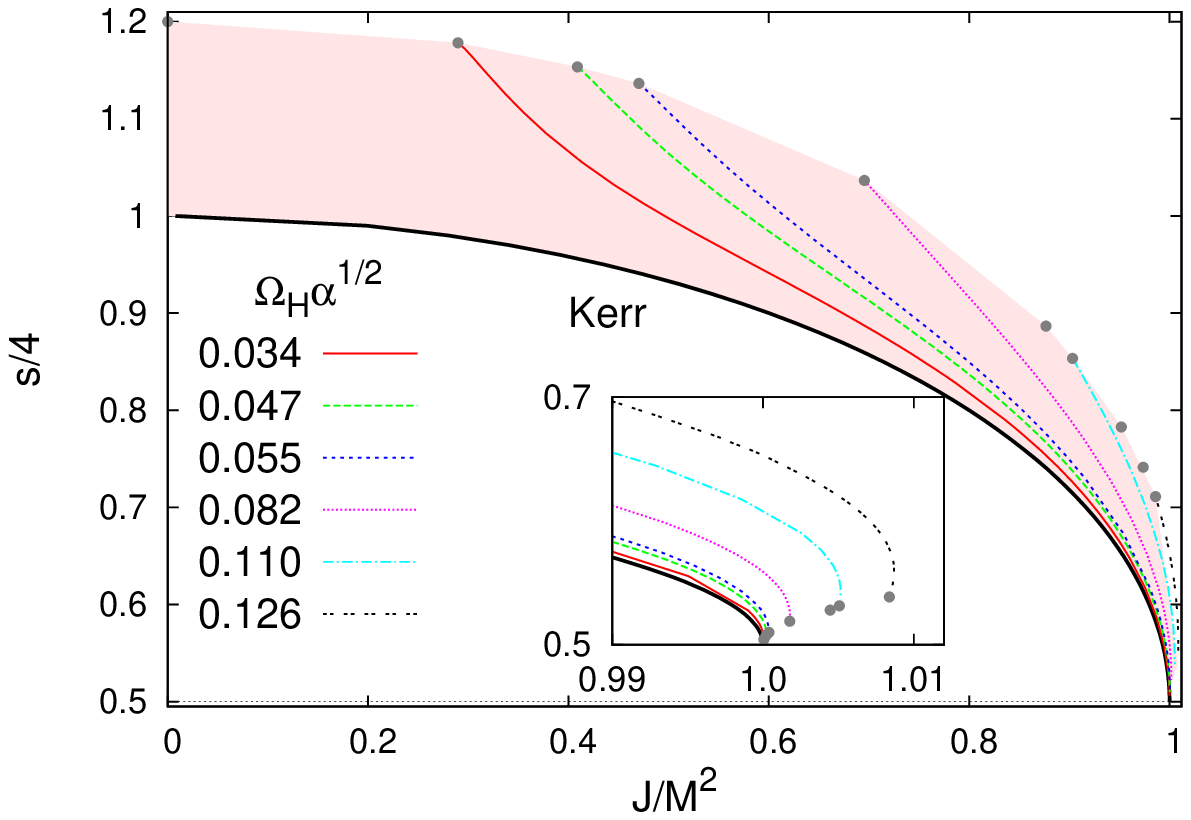}
\includegraphics[height=.25\textheight, angle =0]{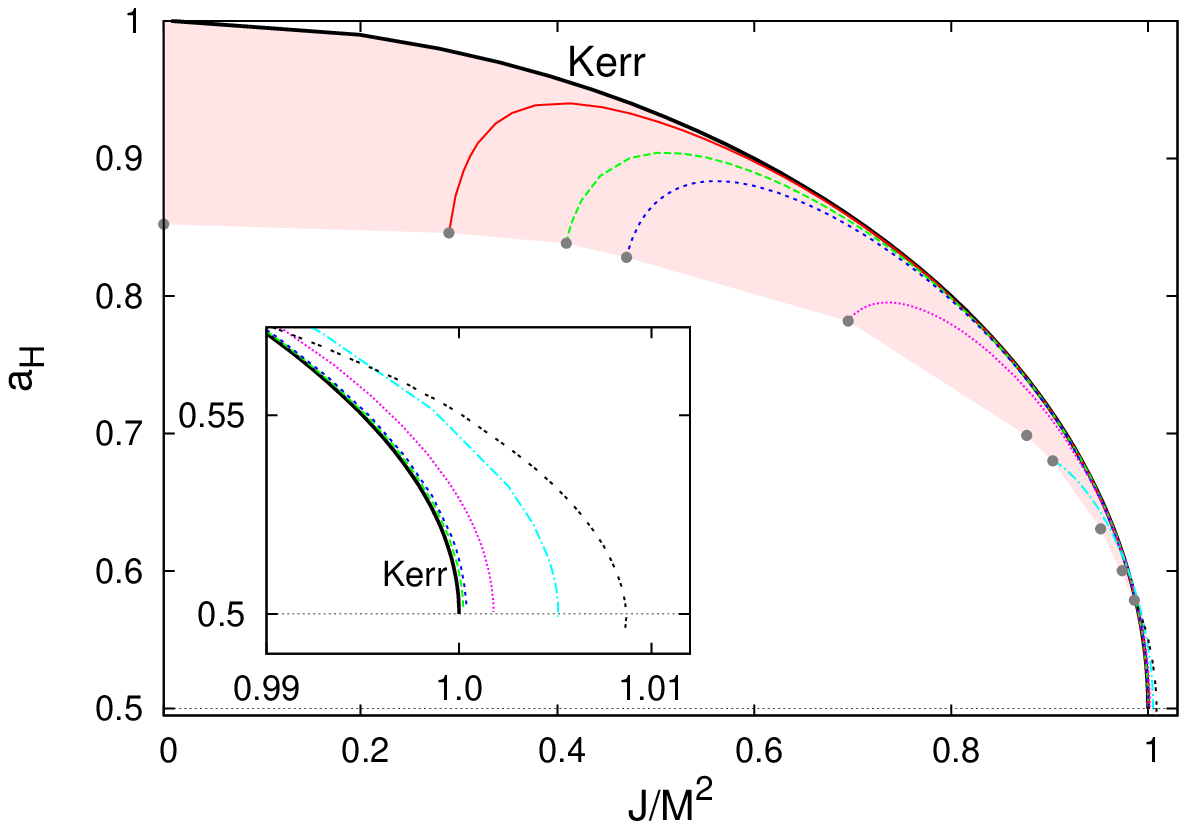}
\end{center}
\caption{
The scaled entropy $S=S/M^2$ (left) 
and scaled horizon area $a_{\rm H}=A_{\rm H}/M^2$ (right)
versus the scaled angular momentum $j=J/M^2$ 
for families of EBGd BHs with fixed scaled horizon angular
velocity $\Omega_{\rm H} \alpha/^{1/2}$.
The shaded area indicates the domain of existence.
The dots in the inset of the left panel correspond to extrapolated values.
}
\end{figure}

Extremal Kerr BHs have vanishing temperature and vanishing
isotropic horizon radius.
Likewise, the families of EBGd BHs end in solutions with
vanishing temperature and vanishing
isotropic horizon radius.
The vanishing of the temperature at the end point of these
curves is seen in Fig.~\ref{f-2} (left), where the scaled 
temperature $T_{\rm H} M$ is exhibited versus
the scaled angular momentum.
Clearly, these endpoints form the lower part of the boundary
of the domain of existence for $j>1$,
while the other boundaries are formed 
by the critical EGBd BHs, the Kerr BHs.

However, unlike the extremal Kerr solution,
the extremal EGBd solutions are not regular.
While the (isotropic) metric functions tend to well defined
limiting functions, with the isotropic horizon radius
and the surface gravity approaching zero,
the dilaton field diverges in this limit at the poles
of the BH horizon, making the extremal solutions singular.

The nonexistence of regular extremal solutions is also seen
when, following $e.g.$ \cite{Astefanesei:2006dd}, one attempts to 
construct the corresponding 
near-horizon geometries with an isometry group $SO(2,1)\times U(1)$.
Again, the dilaton field at the horizon is found to 
diverge at the poles.
Indeed, this holds true independent of the value of the dilaton 
coupling constant, also a perturbative solution exists for small $\alpha$.

At the outermost point of the domain of existence,
where the scaled angular momentum $j$ reaches its maximal value,
the branches of critical EGBd BHs and singular extremal solutions merge
and end. Here the scaled horizon angular velocity reaches
its maximal value, $\Omega_{\rm H} \alpha^{1/2} \approx 0.135-0.14$.

Summarizing the black hole properties, we note that for a  EGBd theory,
specified by the coupling constant  $\alpha$ and $\gamma=1$,
the regular rotating EGBd BHs are found within their domain of existence
delimited by the Kerr BHs, the critical EGBd BHs and the
singular extremal EGBd solutions.
For a given mass and angular momentum,
an EGBd BH has a higher entropy and temperature than a Kerr black hole.
Moreover, EGBd BHs exist beyond the Kerr bound.
The high quality of the numerical solutions is seen from the Smarr relation (\ref{Smarr}),
which is satisfied for the EGBd BH families within an accuracy of $10^{-5}$.

\begin{figure}[t!]
\lbfig{f-2}
\begin{center}
\includegraphics[height=.25\textheight, angle =0]{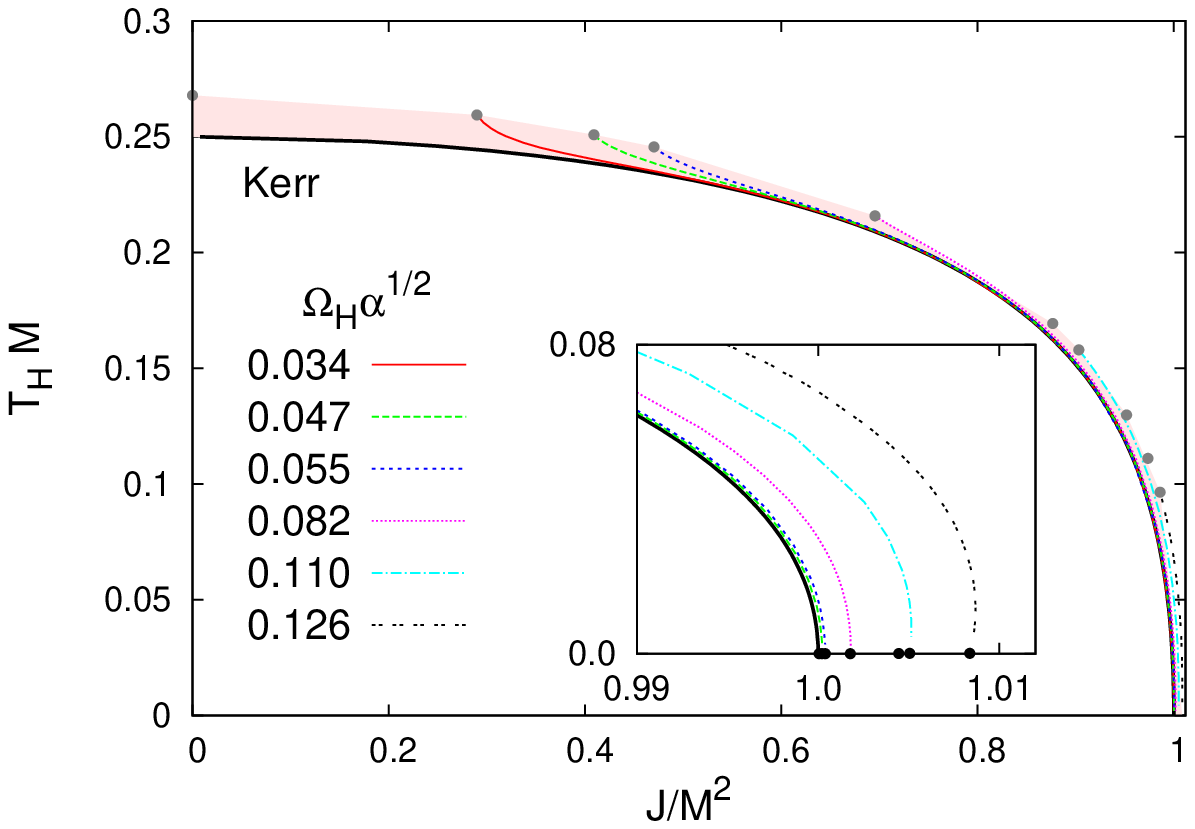}
\includegraphics[height=.25\textheight, angle =0]{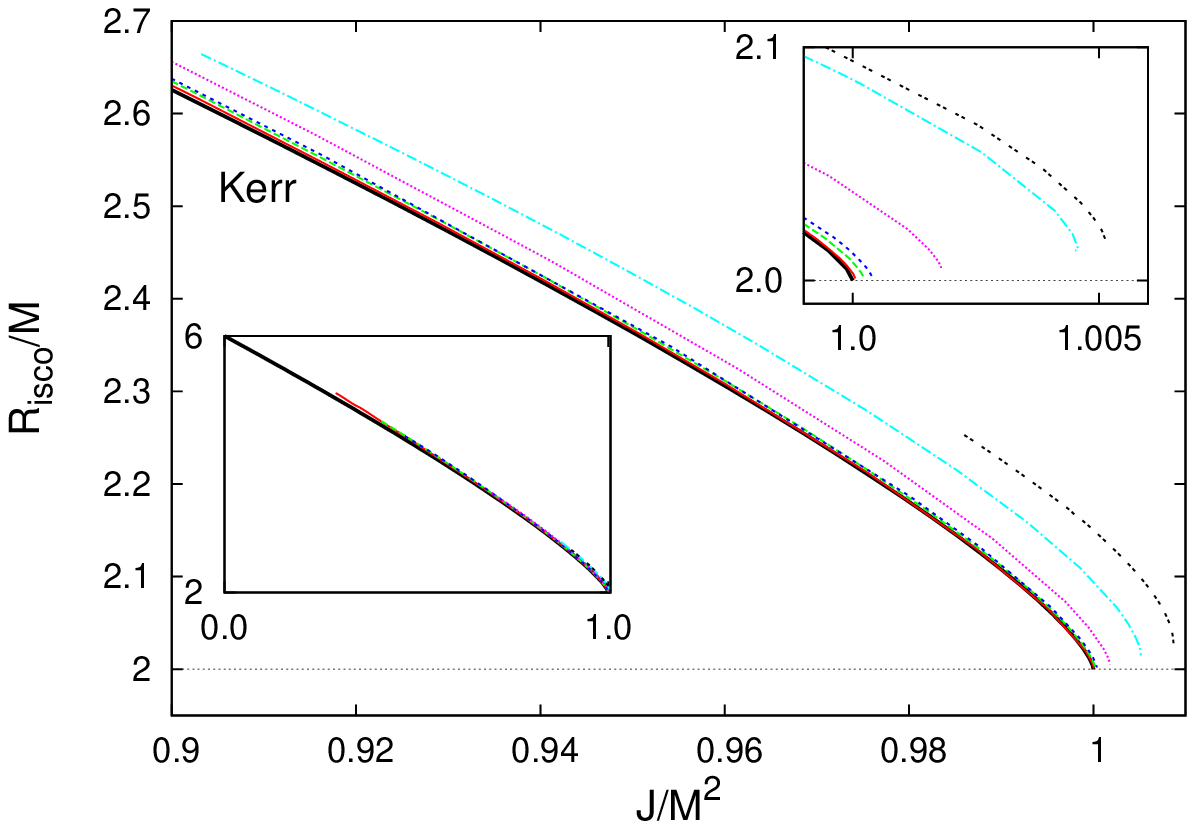}
\end{center}
\caption{
The scaled temperature $T_{\rm H} M$ (left)
and scaled circumferential ISCO radius $R_{\rm ISCO}/M$ 
for dilaton matter coupling constant $\beta=0.5$ (right)
versus the scaled angular momentum $j=J/M^2$
for families of EBGd BHs with fixed scaled horizon angular
velocity $\Omega_{\rm H} \alpha^{1/2}$.
The shaded area indicates the domain of existence.
}
\end{figure}

\noindent{\textbf{~~~Geodesics.--~}}
Let us end by considering possible observational effects
for such EGBd BHs, as addressed before only
for static and slowly rotating solutions \cite{Pani:2009wy}.
For instance,
by studying the geodesic motion around such BHs
the location and orbital frequency of the
innermost-stable-circular-orbit (ISCO) was found to
differ up to a few percent from the Kerr values,
depending on the values of the global charges of the respective solutions
\cite{Pani:2009wy}.

Here we extend this study to the astrophysically relevant regime
of fast rotating BHs. Restricting to motion in the equatorial plane,
we consider timelike geodesics,
with a Lagrangian
$2 {\cal L} = e^{-2 \beta \phi} g_{\mu\nu}\dot x^\mu  \dot x^\nu =-1,$
where the constant $\beta$ fixes the coupling between the matter and the dilaton field 
(with $\beta=0.5$ in heterotic string theory;
also, a superposed dot denotes the derivative with respect to the affine parameter along the geodesics). 
Taking into account the existence of two conserved quantities
associated with the Killing vectors $\xi$ and $\eta$,
one finds the equation $\dot r^2=V(r)$, with the potential $V(r)$ depending on the dilaton and the metric functions
at $\theta=\pi/2$.
 For circular orbits $V(r)=V'(r)=0$; stability of the orbits then requires that the second derivative
of the effective potential is negative. 

We exhibit in Fig.~\ref{f-2} (right)
the scaled circumferential ISCO radius $R_{\rm ISCO}/M$
versus the scaled angular momentum $j=J/M^2$
for families of EBGd BHs with fixed scaled horizon angular
velocity $\Omega_{\rm H} \alpha^{1/2}$,
choosing the dilaton matter coupling $\beta=0.5$.
The maximal values of $R_{\rm ISCO}/M$ correspond to critical EGBd BHs,
while the Kerr values 
and the singular extremal EGBd values form the lower boundary
for the scaled ISCO radius $R_{\rm ISCO}/M$ (for this $\beta$).

For extremal Kerr solutions the ISCO radius is known
to tend to the horizon radius.
Since for extremal Kerr solutions the circumferential horizon
radius corresponds to twice the Boyer-Lindquist radius, 
the scaled ISCO radius $R_{\rm ISCO}/M$ tends to the value two
in the extremal Kerr limit.
For the EGBd solutions we likewise observe,
that the ISCO radius tends to the horizon radius in the 
singular extremal limit.
Here also the dependence on the dilaton coupling $\beta$
vanishes.
In constrast, generically the scaled ISCO radius decreases with decreasing 
$\beta$ towards and (for small $\beta$ and $j$)
below the corresponding Kerr value.
For large scaled angular momentum ($j\le 1$) the deviation
of $R_{\rm ISCO}/M$ from the Kerr value can be as much as 10\%
(for $\beta=0.5$).
Similarly, the orbital frequencies exhibit the largest deviations
from the Kerr frequencies in this range
and can amount to 60\%.
Effects of this size for fast rotating black holes
might be observable in astrophysical systems.

On the more theoretical side we note, that
by including gauge fields further interesting rotating hairy black
holes should be generated, 
representing new solutions of the low energy effective action
of string theory.



\begin{thebibliography}{99}

\bibitem{Mignemi:1992nt}
  S.~Mignemi and N.~R.~Stewart,
  Phys.\ Rev.\  D {\bf 47} (1993) 5259
  [arXiv:hep-th/9212146].
\bibitem{Mignemi:1993ce}
  S.~Mignemi,
  Phys.\ Rev.\  D {\bf 51} (1995) 934
  [arXiv:hep-th/9303102].
\bibitem{Kanti:1995vq}
  P.~Kanti, N.~E.~Mavromatos, J.~Rizos, K.~Tamvakis and E.~Winstanley,
  Phys.\ Rev.\  D {\bf 54} (1996) 5049
  [arXiv:hep-th/9511071].
\bibitem{Torii:1996yi}
  T.~Torii, H.~Yajima and K.~i.~Maeda,
  Phys.\ Rev.\  D {\bf 55} (1997) 739
  [arXiv:gr-qc/9606034].
\bibitem{Alexeev:1996vs}
  S.~O.~Alexeev and M.~V.~Pomazanov,
  Phys.\ Rev.\  D {\bf 55} (1997) 2110
  [arXiv:hep-th/9605106].
\bibitem{McClintock:2006xd}
  J.~E.~McClintock, R.~Shafee, R.~Narayan, R.~A.~Remillard, S.~W.~Davis and L.~X.~Li,
  Astrophys.\ J.\  {\bf 652} (2006) 518
  [arXiv:astro-ph/0606076].
\bibitem{Pani:2009wy}
  P.~Pani and V.~Cardoso,
  Phys.\ Rev.\  D {\bf 79} (2009) 084031
  [arXiv:0902.1569 [gr-qc]].
\bibitem{wald}
 R.~M. Wald, 
 General Relativity 
 (University of Chicago Press, Chicago, 1984)
\bibitem{Kleihaus:2000kg}
  B.~Kleihaus and J.~Kunz,
  Phys.\ Rev.\ Lett.\  {\bf 86}, 3704 (2001)
  [arXiv:gr-qc/0012081].
 \bibitem{Wald:1993nt}
  R.~M.~Wald,
  Phys.\ Rev.\  D {\bf 48} (1993) 3427
  [arXiv:gr-qc/9307038].
\bibitem{Kleihaus:2002tc}
  B.~Kleihaus, J.~Kunz and F.~Navarro-Lerida,
  Phys.\ Rev.\ Lett.\  {\bf 90}, 171101 (2003)
  [arXiv:hep-th/0210197].
\bibitem{Astefanesei:2006dd}
  D.~Astefanesei, K.~Goldstein, R.~P.~Jena, A.~Sen and S.~P.~Trivedi,
  JHEP {\bf 0610} (2006) 058
  [arXiv:hep-th/0606244].


\end{thebibliography}
\end{document}